# Investigating Reliability Aspects of Memristor based RRAM with Reference to Write Voltage and Frequency

T. D. Dongale [a, $], K. V. Khot [b], S. V. Mohite [c], N. K. Desai [a], S. S. Shinde [c], A. V. Moholkar [c], K. Y. Rajpure [c], P. N. Bhosale [b], P. S. Patil [c], P. K. Gaikwad [d], R. K. Kamat [d]

[a] Computational Electronics and Nanoscience Research Laboratory,
School of Nanoscience and Biotechnology, Shivaji University Kolhapur- 416004
[b] Department of Chemistry, Shivaji University, Kolhapur 416004, India
[c] Department of Physics, Shivaji University, Kolhapur 416004, India
[d] Embedded System and VLSI Research Laboratory, Department of Electronics,
Shivaji University, Kolhapur, 416004, India

**Abstract**

In this paper, we report the effect of write voltage and frequency on memristor based Resistive Random Access Memory (RRAM). The above said parameters have been investigated on the linear drift model of memristor. With a variation of write voltage from 0.2V to 1.2V and a subsequent frequency modulation from 1, 2, 4, 10, 100 and 200 Hz the corresponding effects on memory window, Low Resistance State (LRS) and High Resistance State (HRS) have been reported. Thus the lifetime ($\tau$) reliability analysis of memristor based RRAM is carried out using above results. It is found that, the HRS is independent of write voltage, whereas LRS shows dependency on write voltage and frequency. The simulation results showcase that the memristor possess higher memory window and lifetime ($\tau$) in the higher voltage with lower frequency region, which has been attributed to the fewer data losses in the memory architecture.

[$] **Corresponding Author:** T. D. Dongale
E-mail: tukaram.eln@gmail.com

1. Introduction

Traditionally the passive circuit family in Electrical Engineering consists of three lumped elements viz. resistor, inductor and capacitor. In 2008, HP research group reported the first physical realization of fourth fundamental circuit element named as *Memristor* [1], which was postulated mathematically much before four decades by Prof. L. Chua [2]. Memristor is a nonlinear circuit element and possess nonvolatile memory property, which is not observed in any other circuit elements. The inherent memory property of memristor is distinctly observed in the nanoscale and therefore it is considered as a strong candidate for next generation memories [3]. Along with its applications in memory, there are many interesting applications explored around memristor such as neuromorphic computations [4], in-memory computing [5], biomedical appliances [6] and many more.

Recently, Nickel et al. developed the nonlinear, bipolar memristor crossbar structures and demonstrated the high scalability in the developed device [7]. Emara et al. reported the 1T2M differential memory cell for single and multi bit data storage [8]. Ning et al. proposed the nonvolatile threshold adaptive transistors model with embedded RRAM for neuromorphic application [9]. Dongale et al. reported the $TiO_2$ thin film memristor with low symmetric voltage switching [10]. Hoessbacher et al. reported a novel application of memristor in plasmonic domain. The reported plasmonic memristor can be used as electrically activated optical switches with a memory effect [11]. Murali et al. reported the zinc–tin-oxide (ZTO) based memristor. Good switching ratio, long retention time, and good endurance have been observed in the developed memory device [12]. In the backdrop of the international research scenario, our research group is also striving hard to model and realize memristor using different methods and come out with useful applications [13-16]. Previously we have reported the effect of device size as a function of frequency on memristor based RRAM [17]. We have also investigated the conduction mechanism and frequency dependency of nanostructured memristor device [18-19].

In the present paper, we have reported the effect of write voltage as a function of frequency on memristor based Resistive Random Access Memory (RRAM) and thereby endured reliability of the device. The proposed investigation is based on linear drift model of memristor proposed by HP research group [1]. The rest of the paper is as follows, after a brief introduction in the first section, the second section deals with the computational details of simulation, followed by effect of write voltage as a function of frequency on memristor based RRAM in the third section. The fourth section investigates the effect of write voltage as a function of frequency variation on lifetime (τ) reliability of memristor device. At the end conclusion is portrayed.

## 2. Computational Details

The memristor properties are simulated using linear drift model of the device considering the Pt/TiO$_2$/Pt structure [1]. For the present investigation drift velocity of oxygen vacancies is considered as a state variable 'w'. The details of conduction mechanism and device structure has been investigated thoroughly by many researchers [14-15, 20-22]. Considering the linear ionic drift with the average drift velocity of oxygen vacancies $\mu_V$, leads us to memristor current and voltage relation represented by following mathematical equations: [1]

$$V(t) = \left[\left(\frac{R_{ON}\ W(t)}{D}\right) + R_{OFF}\left(1 - \frac{W(t)}{D}\right)\right] i(t) \quad \text{------- (1)}$$

Where state variable 'w' can be represented as,

$$\frac{dw(t)}{dt} = \eta\ \frac{\mu_v\ R_{ON}}{D}\ i(t) \quad \text{------- (2)}$$

where, *V(t)* represents applied voltage, *i(t)* is the current through device, *D* is the device size, $R_{ON}$ and $R_{OFF}$ considered as LRS and HRS of the device respectively and $\eta$ represents the polarity of the device. The present simulation is carried out for 10 nm memristor device in which doped region is about 2 nm and drift velocity of oxygen vacancies is $\mu_V=10^{-14} m^2 V^{-1} s^{-1}$. The M-efficiency ($R_{OFF}/R_{ON}$) factor is considered as 200. The amplitude is progressively increased in the step of 0.2 V and corresponding effect has been observed on LRS and HRS.

## 3. Effect of Write Voltage and Frequency on Memristor based RRAM

The write voltage plays a key role in the resistive switching of memristor device. The properties of memristor are distinctly observed in the nanoscale regions and at this region small bias can produce large electric field across the device which accelerates the drift velocity of charge carriers. This large electric field produces nonlinear drifting of vacancies near the boundary interfaces [20-22].

In view of this the fig. 1 (a) represent the pinched hysteresis loop (PHL) at 0.2 V and remaining results are for 0.4 V, 0.6 V, 0.8 V. 1 V and 1.2 V (fig. b to f) respectively. The bipolar resistive switching is observed for each case. The areas under bipolar resistive switching (pinched hysteresis loop) seem to be increasing with the corresponding increase in write voltage i.e. from 0.2 V to 1.0 V. After 1V the loop area tends to decrease. Generally, the PHL is considered as a memory property of memristor. The results clearly show that memristor requires appropriate bias for bipolar resistive switching. In the present investigation, the highest bipolar resistive switching is of the order of 0.8 V – 1.0 V. This typical loop area of memristor is termed as *Memristance* and the value thereof is one of the performance metrics of Resistive Random Access Memory (RRAM).

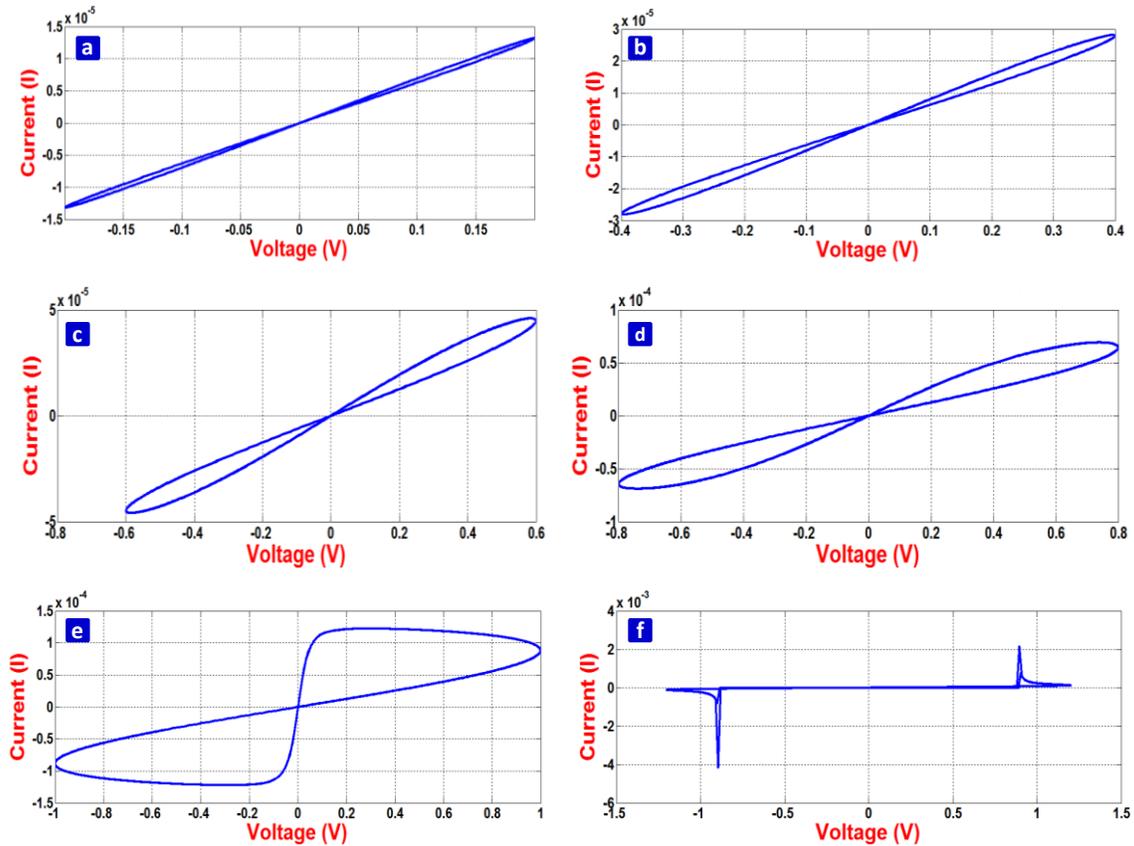

**Fig.1 (a-f):** Pinched Hysteresis Loop (PHL) of Memristor for Different Write Voltages. Fig. (a) Represents the PHL for 0.2 V and the remaining results are for 0.4 V, 0.6 V, 0.8 V. 1 V and 1.2 V (Fig. b to f) respectively.

The increase in the memristance is due to the fact that the concentration and motilities of vacancies are increased in the range of particular bias [21-23]. The loop area is degraded above 1V and shows highly nonlinear behavior at 1.2 V. The nonlinear behavior at this range is attributed to large electric field across the 10 nm device. This large electric field produces nonlinear drifting of vacancies near the boundary interfaces, which further results in high nonlinearity in the I-V characteristics.

The results indicate that, 1V is an upper threshold for memristor when it is used for RRAM applications at 10 nm device dimension. The change in the bias also alters the response of the charge-magnetic flux plot. The charge-magnetic flux plot is linear at the lower value of write voltage (0.2V-0.4V) and it becomes nonlinear at higher value of write voltage (0.6V-1.2V), which is shown in the fig. 2(a-f). The symmetric nature of I-V plot is also gets altered when bias of the device is changed, which is shown in the fig. 3(a-f). These results indicate that the memristor possess bias dependent resistive switching property.

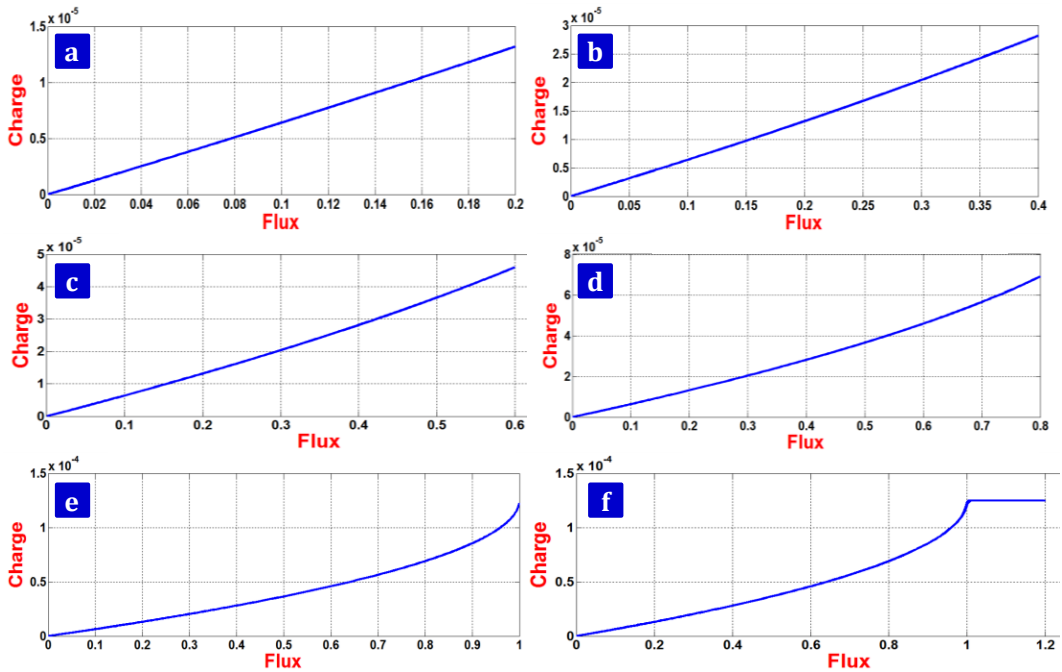

**Fig.2 (a-f):** Effect of write voltage on Charge (q) - Flux ($\varphi$) relation.

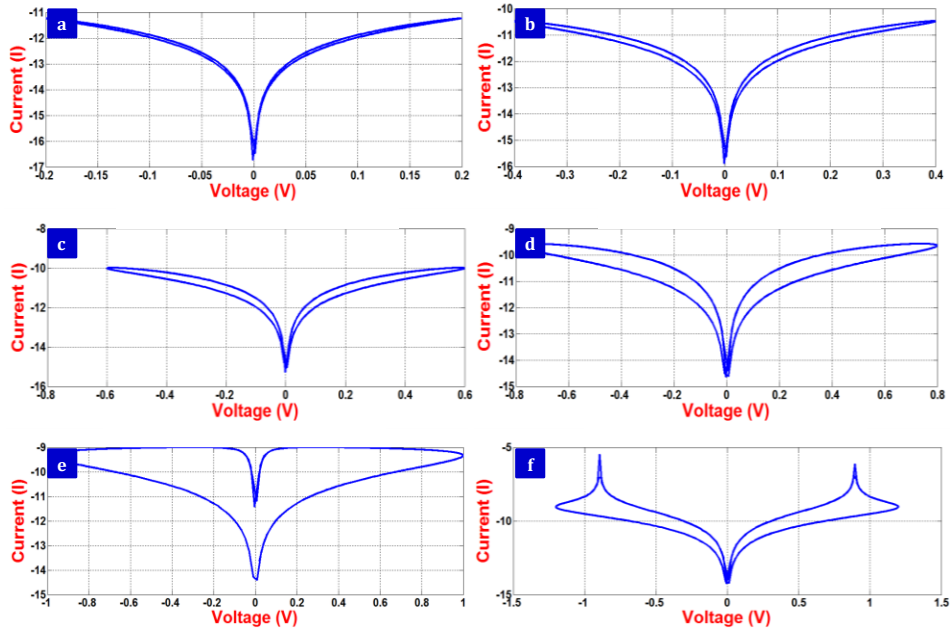

**Fig.3 (a-f):** Effect of write voltage on semi-log Current (I) - Voltage (V) property.

The results of present simulations suggest that the current multiplier factor plays an important role for resistive switching and is found to be $10^{-5}$ A for 0.2 V to 0.6 V. The same becomes $10^{-4}$ A, $10^{-3}$ A for the 0.8V to1.0 V, and 1.2 V respectively. In the nano scale, current threshold for resistive switching seems to be operational and it further evidences that the rate of drifting of oxygen vacancies (state variable of Memristor) is higher at particular range of bias and lower for other biases. The simulation results clearly evident that memristor is a very apt fundamental building block for the nonvolatile memory with high degree of symmetry. The low threshold bias switching in the memristor makes it a promising candidate for future low power consumption memories.

The effect of write voltage on LRS and HRS is also shown in fig. 4 and 5. It is found that the LRS decreases as the applied write voltage increases and there is no effect observed on HRS. The memory window is high only at lower frequencies (1 Hz and 2 Hz) and tends to diminish at higher frequencies. The very small change in the LRS is observed at higher frequencies. In actual practice memory window should be high to detect the two resistance states easily by read/write circuits. The memory window is one of the performance metrics of high performance RRAM and it is associated with correction of data reading. The small memory window leads to read/write errors in the memory architecture [17, 24]. The results associated with the LRS are in good agreement with the result reported by Nili et al. [25] However the HRS results are not getting along with the above. This is attributed to the different physical mechanisms getting associated with the each memristor device.

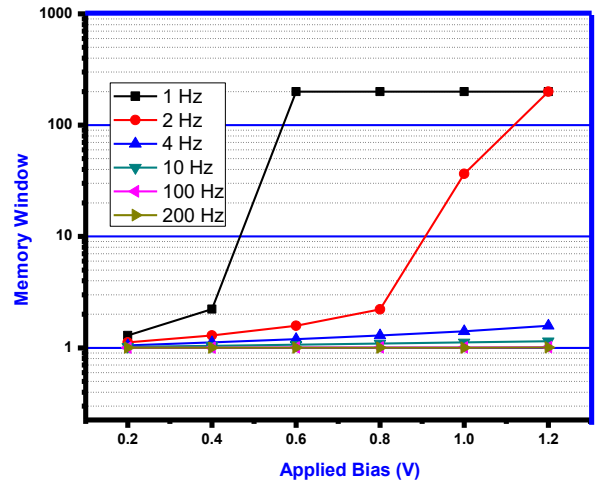

**Fig.4:** The Effect of Write Voltage on Memory Window of Memristor with Change in Frequency. The results clearly indicate that the memory window decreases as the frequency of applied signal is increases.

Thus the results clearly depict existence of higher memory window only at higher write voltage with lower frequency. This results into fewer read/write errors in the memristor based RRAM. The results also suggested that the memory window becomes very small at higher frequency hence read/write errors becomes dominant and hence memristor based RRAM cannot be used in this region.

## 4. Effect of Bias or Write Voltage and Frequency on Memristor Lifetime (τ) Reliability

The data handling capacity of a memory device in general is characterized by its data retention property. The data is supposed to be retained for very long period of time and it is one of the primary requirements for any memory device. The external or internal malfunction or faults can be responsible for data losses [17, 24]. Furthermore this fault also affects the lifetime (τ) reliability of memory device. The lifetime of memristor as well as data retention can be increased by taking proper care of faults and malfunctions. In this backdrop, we have investigated the effect of write voltage and frequency on reliability of memristor based RRAM. The results described in the previous sections confirm the important role of write voltage in the memristor based RRAM. With reference to the write voltage scenario, the value of LRS is very low for lower voltage region, and it becomes high for higher voltage region. At the same time, the value of LRS increases as the frequency of applied signal increases which is shown in the fig. 5.

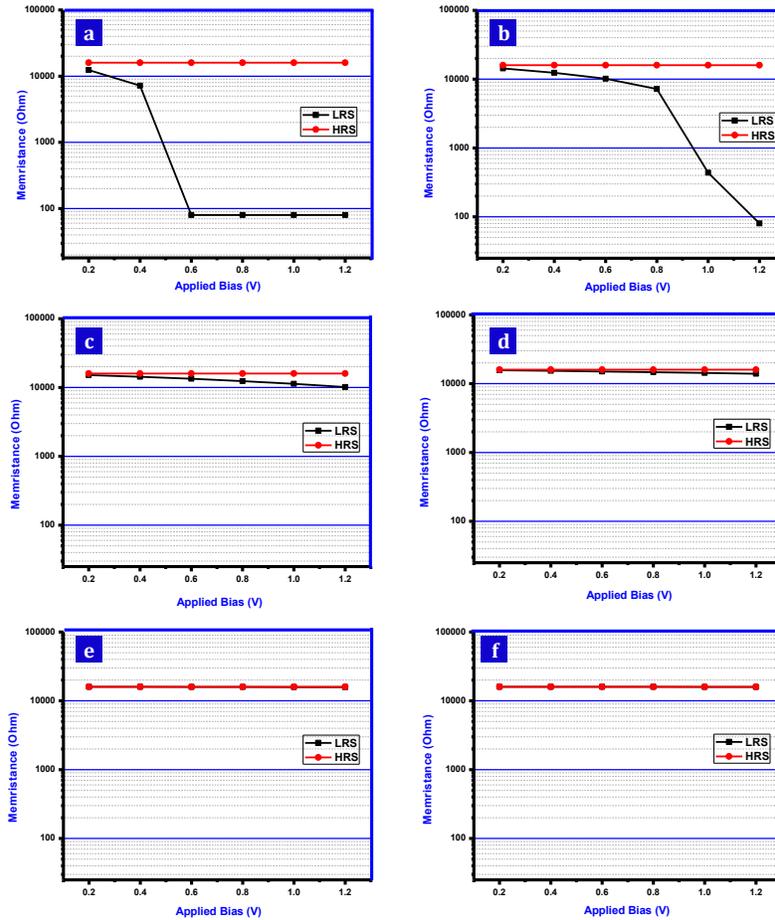

**Fig. 5 (a-f):** The Effect of Write Voltage on Low Resistance State (LRS) and High Resistance State (HRS) with change in Frequency. Fig. (a) represents, the change in LRS and HRS with change in write voltage at 1 Hz. Fig. (b to f) represents the change in LRS and HRS with change in frequency at 2, 4, 10, 100, and 200 Hz respectively. The results clearly indicate that the LRS is decreases for low frequencies up to 10 Hz and HRS remains same for all voltage and frequencies. The memory window is higher at high voltage and low frequency of signal.

The low value of LRS is also responsible for higher peak current. If peak current is gets decreased by any malfunction or process variation then lifetime of memristor based RRAM too worsens. If the peak current is reduced to its minimum threshold value, then sense amplifier cannot distinguish the difference between data and noise. This further leads to the data loss in the memory architecture [17, 24]. The results of our investigation clearly show that the memristor possess higher lifetime (τ) in the higher voltage region with lower frequency of applied signal, which is shown in the fig. 6. It is due to the fact that, the peak current is higher for high voltage region with lower value of LRS and frequency. It is also seen that the lifetime (τ) is degraded at higher frequencies. The lower voltage region of RRAM shows the lower value of peak current; hence sense amplifier does not distinguish between data and noise. Therefore any process variations which leads to lower peak current values is responsible for the lower lifetime (τ) of memristor based RRAM. The results also confirm that the higher bias with lower frequency region could be the best possible domain to get higher lifetime (τ) of memristor based RRAM.

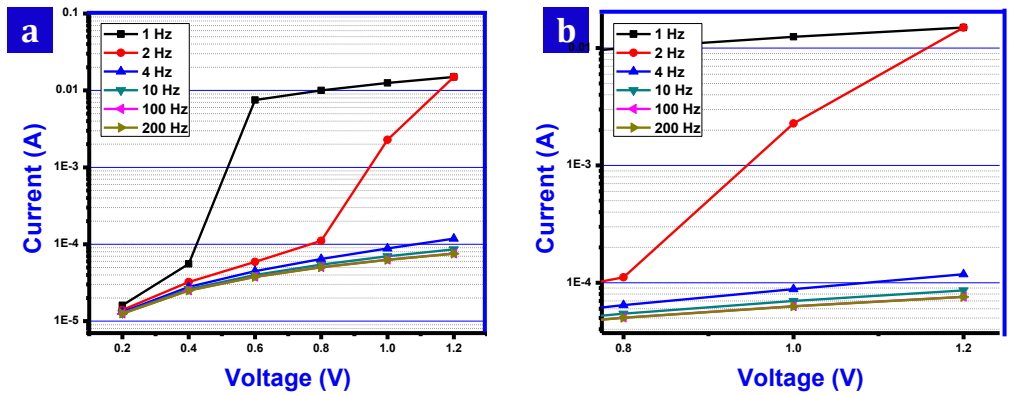

**Fig. 6: (a)** Effect of Write Voltage on Memristor Based RRAM. The current is increased as the Write Voltage increases from 0.2 V to 1.2 V. **(b)** Zoomed view of current at 0.8 V to 1.2 V.

## 5. Conclusions:

The present paper portrayed the effect of write voltage and frequency on memristor based Resistive Random Access Memory (RRAM). It is found that the LRS is a function of write voltage and frequency but HRS is independent of write voltage and frequency. It is further revealed that memory window tends to increase with increase in the write voltage. Furthermore, the memory window is found to be the function of frequencies and higher memory window is achieved at lower frequencies. The lifetime (τ) reliability analysis of memristor based RRAM is carried out using LRS results. It is observed that memristor possess higher lifetime (τ) in the higher voltage with lower frequency region, which results in the lower data losses in the memory architecture.